\documentclass{pasj00}
\draft

\begin{document}
\SetRunningHead{S. Takita et al.}{AKARI IRC slow-scan observations}

\title{Slow-scan Observations with the Infrared Camera (IRC) on-board AKARI.}

\author{
  Satoshi   {\sc Takita}, \altaffilmark{1}
  Norio     {\sc Ikeda}, \altaffilmark{1}
  Yoshimi   {\sc Kitamura}, \altaffilmark{1}
  Daisuke   {\sc Ishihara}, \altaffilmark{2}
  Hirokazu  {\sc Kataza}, \altaffilmark{1}
  Akiko     {\sc Kawamura}, \altaffilmark{3}
  Shinki    {\sc Oyabu}, \altaffilmark{2}
  Munetaka  {\sc Ueno}, \altaffilmark{1}
  \and
  Issei     {\sc Yamamura}, \altaffilmark{1}
}
\email{takita@ir.isas.jaxa.jp}

\altaffiltext{1}{Institute of Space and Astronautical Science, 
  Japan Aerospace Exploration Agency, 
  3-1-1 Yoshinodai, Chuo, Sagamihara,
  Kanagawa 252-5210, Japan}
\altaffiltext{2}{Graduate School of Science, Nagoya University,
  Furo-cho, Chikusa-ku, Nagoya, Aichi 464-8602, Japan}
\altaffiltext{3}{National Astronomical Observatory of Japan,
  2-21-1 Osawa, Mitaka, Tokyo 181-8588, Japan}


%

\KeyWords{infrared: general
--- techniques: image processing
--- space vehicles
} 

\maketitle

\begin{abstract}
We present the characterization and calibration of the slow-scan observation mode of the Infrared Camera (IRC) on-board AKARI.
The IRC slow-scan observations were operated at the \textit{S9W} (9~$\mu$m) and \textit{L18W} (18~$\mu$m) bands.
We have developed a toolkit for data reduction of the IRC slow-scan observations.
We introduced a ``self-pointing reconstruction'' method to improve the positional accuracy to as good as $1''$.
The sizes of the point spread functions were derived to be $\sim6''$ at the \textit{S9W} band  and $\sim7''$ at the \textit{L18W} bands in full width at half maximum.
The flux calibrations were achieved with the observations of 3 and 4 infrared standard stars at the \textit{S9W} and \textit{L18W} bands, respectively.
The flux uncertainties are estimated to be better than 20~\% from comparisons with the AKARI IRC PSC and the WISE preliminary catalog.

\end{abstract}

\section{Introduction}

AKARI is the first Japanese infrared astronomical satellite \citep{murakami2007}, which was launched on 2006 February 21.
AKARI was brought into a sun-synchronous polar orbit at an altitude of 700~km.
AKARI has a 68.5~cm diameter cooled telescope with the two scientific instruments of the Infrared Camera (IRC; \cite{onaka2007}) for 1.8--26~$\mu$m and the Far-Infrared Surveyor (FIS; \cite{kawada2007}) for 50--180~$\mu$m.
One of the major observation programs of AKARI is an all-sky survey at the mid- to far-infrared wavelengths with 6 photometric bands to provide second generation infrared catalogs.
In addition, AKARI has the capability for imaging and spectroscopy in the pointed observation mode.

The IRC is mainly designed for deep imaging and spectroscopy over 1.8--26~$\mu$m with three independent channels:
NIR (1.8--5.5~$\mu$m), MIR-S (6--12~$\mu$m), and MIR-L (12--26~$\mu$m).
All the channels have filter wheels, which hold three filters and two spectroscopic dispersers.
The field-of-view (FoV) is $10' \times 10'$.
The MIR-S and MIR-L channels have infrared sensor arrays of 256~$\times$~256~pixels.
The pixel sizes for MIR-S and MIR-L are 2\farcs34~$\times$~2\farcs34 and 2\farcs51~$\times$~2\farcs39, respectively.
The FoVs of MIR-S and MIR-L are separated by $20'$ perpendicularly to the AKARI's scan direction.

Although IRC was originally designed for imaging and spectroscopy in the pointed observation mode, \citet{ishihara2006} developed an additional function, ``scan mode'', to carry out the All-Sky Survey with the MIR-S and MIR-L channels.
In the ``scan mode'', we used 2 out of 256 rows in the sensor array for continuous and non-destructive readout.
The data were sampled at every 0.044 sec, and the pixels were reset at every 13.5 sec (or every 306 samplings) to discharge the photo-current.
The output signals of every four adjacent pixels along each row were binned on board to reduce the data amount, so the effective pixel size in the cross-scan direction was about 10$''$.
The details of the IRC ``scan mode'' and the all-sky survey are described in \citet{ishihara2006} and \citet{ishihara2010}.

\section{IRC Slow-scan Observations}

The IRC slow-scan observation mode was designed for large-area (up to $10' \times 1\fdg5$ in one pointed observation) mapping observations with sufficient sensitivities, using the ``scan mode''.
The IRC slow-scan observations are provided in the Astronomical Observation Templates (AOTs) of IRC11 and IRC51.
In the slow-scan observations, the telescope scanned along one or two round trip paths around the target object to take a redundant dataset.
More details of the telescope operations in the slow-scan observations are described in \citet{kawada2007}\footnote{IRC and FIS were simultaneously operated in the slow-scan mode.}.

The basic properties of the slow-scan mode are the same as in the All-Sky Survey:
(1) the data of the \textit{S9W} (9~$\mu$m) and \textit{L18W} (18~$\mu$m) bands were obtained with the MIR-S and MIR-L channels, respectively
\footnote{Only one set of observations was carried out with the \textit{S11} (11~$\mu$m) band with the MIR-S channel \citep{ishihara2007}.},
(2) the data of only 2 rows in the 256 $\times$ 256 pixel array in each channel were acquired, and
(3) the data were sampled at every 0.044 sec with non-destructive readout.
On the other hand, there are three points different from those in the All-Sky Survey as follows:
\begin{enumerate}
\item The pixels were reset at every 306 and 51 samplings to discharge the photo-current for IRC11 and IRC51, respectively.
\item The telescope scanned the sky with a much slower speed (8, 15 or 30$''$ sec$^{-1}$) than that of the All-Sky Survey ($216''$ sec$^{-1}$).
\item The output signals of every four adjacent pixels along each row were binned on board, so that the effective pixel size in the cross-scan direction is 10$''$ for IRC11 (same as the All-Sky Survey), while no binning was applied for IRC51 so that the pixel size is 2\farcs5.
\end{enumerate}

Since the array operations of IRC11 are identical with the All-Sky Survey and most of the slow-scan observations were performed with IRC51, we describe IRC51 observations in this paper.

\section{Data Packages of the IRC Slow-scan Mode}

Data are stored in dedicated format FITS file, called Time-Series Data (TSD), which was originally developed for the FIS (see \cite{fis_dum}), for the data package of the IRC slow-scan observations to handle the data easily.
The TSD is a binary FITS table and consists of a header part and arrays of data records.
One record consists of the IRC instrument raw data and necessary information from other house keeping (HK) instruments, as well as positional information given by the ground attitude determination processing.
Since the sampling rates of these data are different from each other, the data should be interpolated to synchronize exactly with the clock of the detectors.
Two data sets for the MIR-S and MIR-L channels of the IRC are created from one pointed observation.

\section{Data Reduction Processes}
\label{sc:aris}

We have developed a dedicated software for reduction of the IRC slow-scan data, ARIS (AKARI data Reduction tools for the IRC Slow-scan).
ARIS is written in Interactive Data Language (IDL).
Most of the ARIS processes handle the TSD format data.
We describe individual data reduction processes in the following sub-sections.

\subsection{Basic calibrations}
In the first step of data processing, we applied the following basic calibrations.
First, we corrected the anomalous behaviour of the detector output, which is seen after the reset.
Second, we corrected the non-linearity between incoming photons and output signals.
Third, we differentiated the data, because the output signals from the detectors are time integral values.
Fourth, we subtracted the dark signals.
Then, we applied the flat fielding.
Finally, we masked bad data, such as pixels which were masked out for the slit spectroscopy, pixels just after the resets, and saturated pixels.

\subsubsection{Reset anomaly correction}
\begin{figure}
\begin{center}
\FigureFile(80mm,55mm){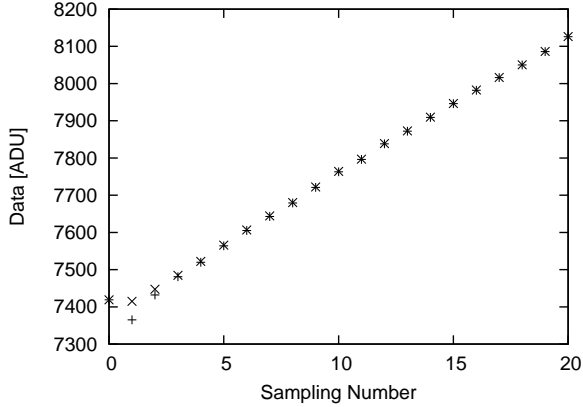}
\end{center}
\caption{Example of the reset anomaly correction.
The plus and cross symbols are the output signals of a pixel from observed sky and corrected data, respectively.
}
\label{fg:reset}
\end{figure}

There is an anomalous behaviour of the detector output which persists for a few seconds after every reset (the reset anomaly).
Figure \ref{fg:reset} shows an example of the anomaly.
Since the offset level of the detector output is sensitive to the temperature \citep{ishihara2003}, the anomaly can be well described by the temperature drift caused by the tiny reset current to discharge the stacked photo-electrons.
The reset anomaly is corrected as
\begin{equation}
S^{(1)}_i(t) = S^{(0)}_i(t) + R(t),
\label{eq:resetanom}
\end{equation}
where $S^{(0)}$, $S^{(1)}$, and $R$ are the raw data, the corrected data, and the correction offset, respectively.
The suffix $i$ indicates the pixel number and $t$ is the sampling number from the last reset.
The offset is given by
\begin{equation}
R(t) = a_0 \times \exp(a_1 \times t) + a_2 \times \exp(a_3 \times t^2).
\label{eq:resetparam}
\end{equation}
The parameters we adopted are listed in Table \ref{tbl:reset}.

\begin{table}
\caption{Parameters of the reset anomaly correction.}
\label{tbl:reset}
\begin{center}
\begin{tabular}{crr}
\hline
& MIR-S & MIR-L\\
\hline
$a_0$ & $ 3.098$  & $ 2.721$ \\
$a_1$ & $-0.1045$ & $-0.08774$ \\
$a_2$ & $72.31$   & $ 4.973$ \\
$a_3$ & $-0.4321$ & $-0.1087$ \\
\hline
\end{tabular}
\end{center}
\end{table}

\subsubsection{Non-linearity correction}
\begin{figure*}
\begin{center}
\FigureFile(80mm,55mm){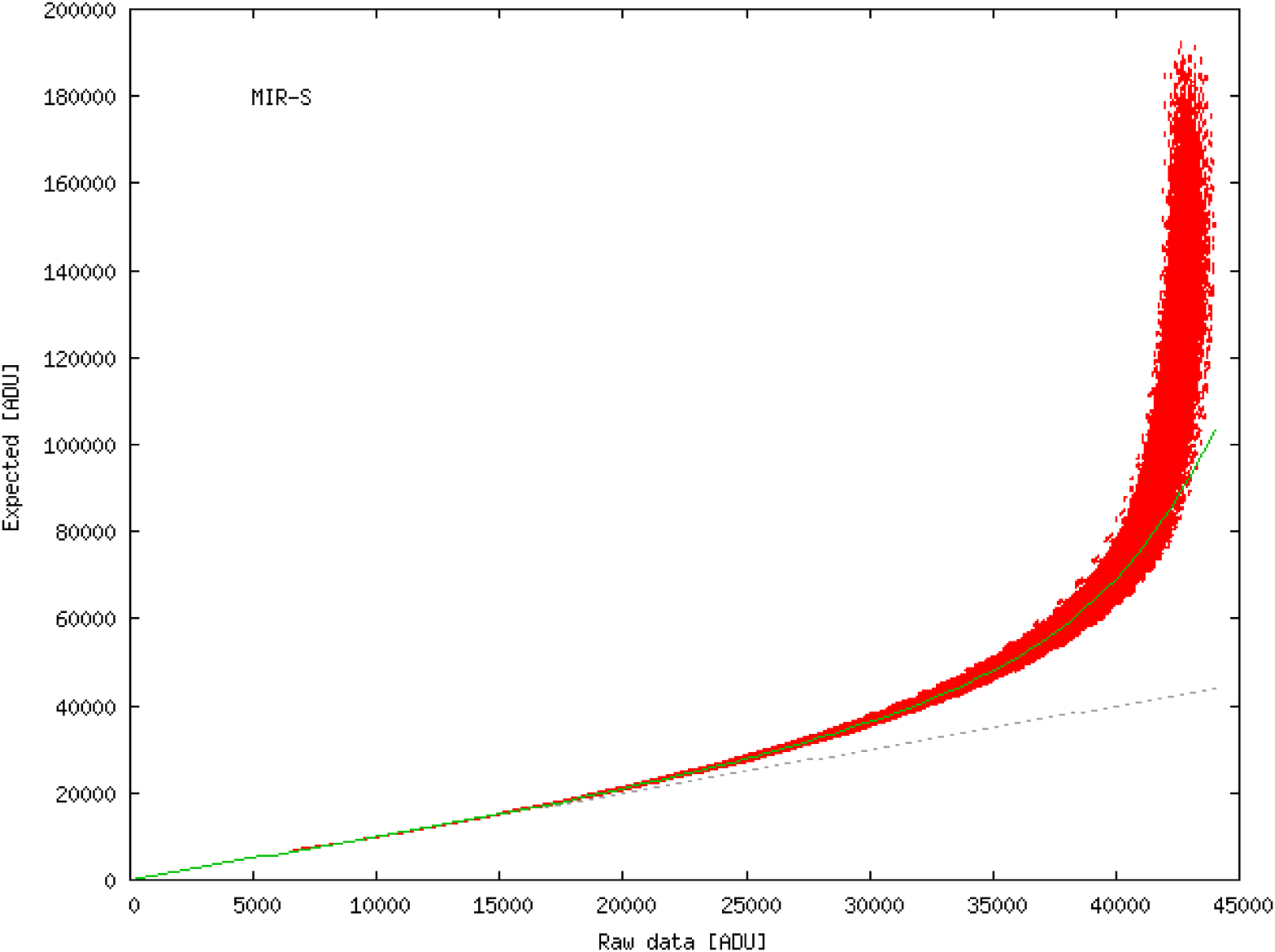}
\FigureFile(80mm,55mm){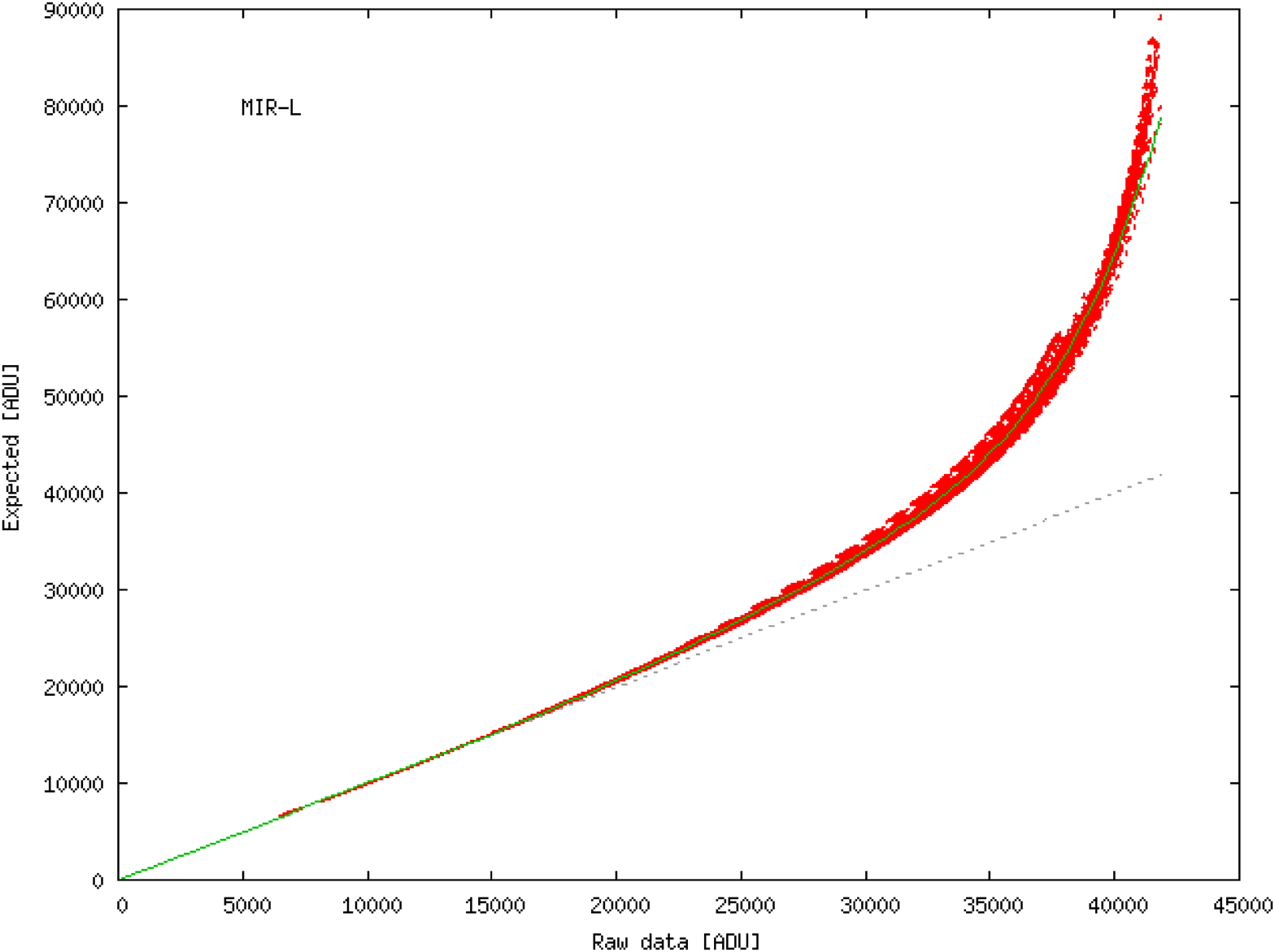}
\end{center}
\caption{Plots of raw signal versus expected signal for the MIR-S (\textit{left}) and MIR-L (\textit{right}) channels.
The green curves are the best-fit polynomial functions given by equation (\ref{eq:linearity}).
The broken lines indicate the best-fit linear function for ADU$<$15000.
}
\label{fg:linearity}
\end{figure*}

\begin{figure*}
\begin{center}
\FigureFile(80mm,55mm){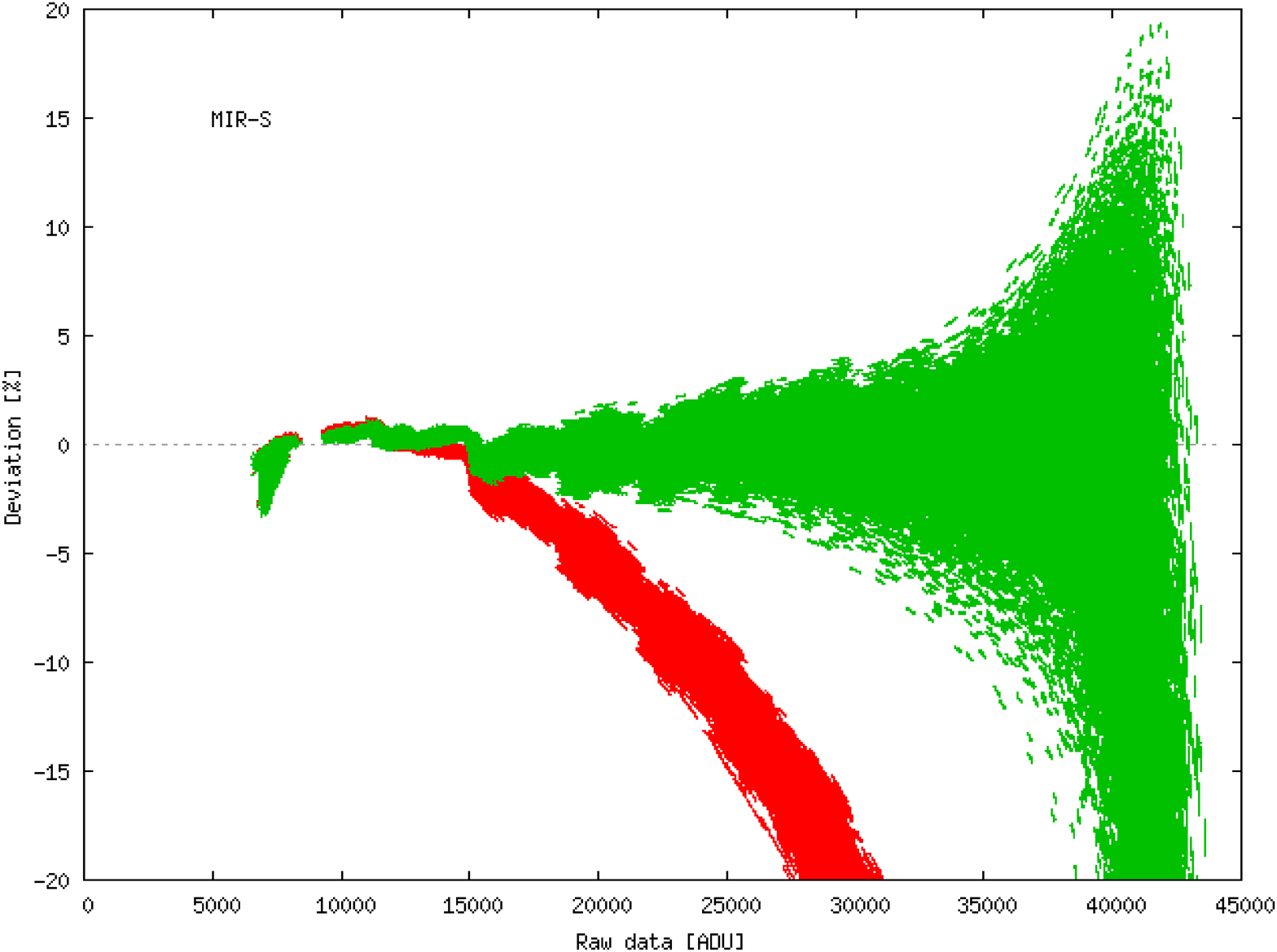}
\FigureFile(80mm,55mm){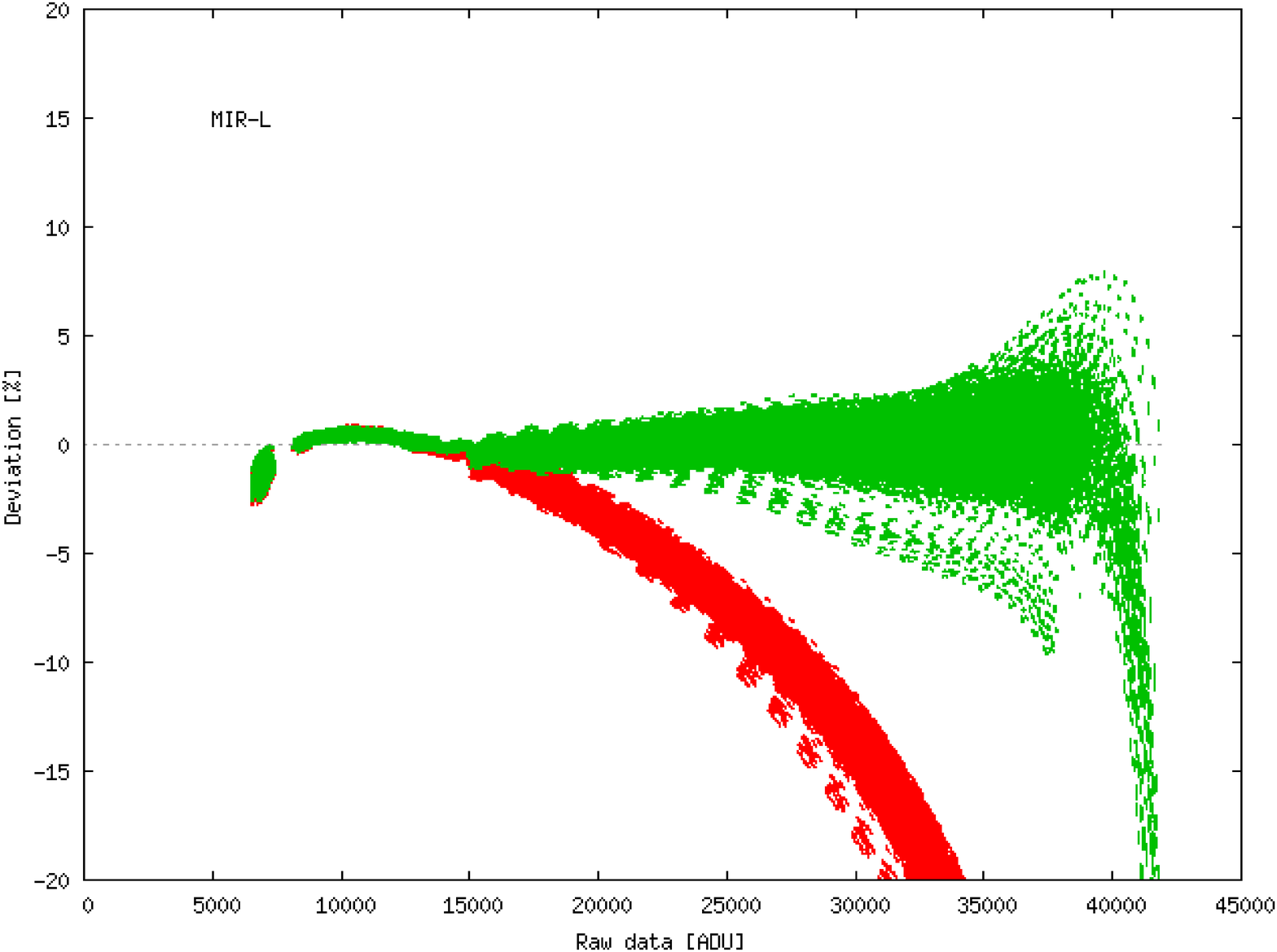}
\end{center}
\caption{Deviations from the best-fit polynomial functions (green dots) for the MIR-S (\textit{left}) and MIR-L (\textit{right}) channels.
The red dots are the deviations from the best-fit linear functions for ADU$<$15000.
}
\label{fg:lindev}
\end{figure*}

The IRC detectors have non-linearity due to the decrease of the bias voltage during integration.
This non-linearity was measured in the laboratory prior to the launch.
Figure \ref{fg:linearity} shows examples of raw signal output vs expected signal, where the expected signal was derived by fitting a linear function to the raw data of ADU $<$ 15000 with a linear function.
We adopt the correction factors in a polynomial form as
\begin{equation}
S^{(2)}_i = \sum_{x = 1}^n L_x \times (S^{(1)}_i)^x,
\label{eq:linearity}
\end{equation}
where $L_x$ is the coefficient for the term of degree $x$.
We adopt $n=7$ for both the MIR-S and MIR-L channels, and the best-fit parameters are listed in Table \ref{tbl:linearity}.
Figure \ref{fg:lindev} shows the deviations from the best-fit polynomial functions, indicating the polynomial fitting is as accurate as 5~\% for the output signals less than 30000 and 35000~ADU for the MIR-S and MIR-L channels, respectively.
Figure \ref{fg:lindev} also shows that the physical detector saturation occurs around 40000~ADU for both the channels.

\begin{table}
\caption{Best-fit parameters of the non-linearity correction.}
\label{tbl:linearity}
\begin{center}
\begin{tabular}{crr}
\hline
& MIR-S & MIR-L\\
\hline
$L_1$ & $ 0.9977$                & $ 0.9996$ \\
$L_2$ & $ 2.559 \times 10^{-6}$  & $-2.906 \times 10^{-7}$ \\
$L_3$ & $-4.029 \times 10^{-10}$ & $ 5.278 \times 10^{-10}$ \\
$L_4$ & $-7.599 \times 10^{-15}$ & $-1.132 \times 10^{-13}$ \\
$L_5$ & $ 3.481 \times 10^{-18}$ & $ 8.722 \times 10^{-18}$ \\
$L_6$ & $-1.387 \times 10^{-22}$ & $-2.659 \times 10^{-22}$ \\
$L_7$ & $ 1.722 \times 10^{-27}$ & $ 2.913 \times 10^{-27}$ \\
\hline
\end{tabular}
\end{center}
\end{table}

\subsubsection{Differentiation}
Since the raw signal from the detector in scan observations are time-integrated data, we differentiate it with respect to sampling number $t$ as
\begin{equation}
S^{(3)}_i(t) = S^{(2)}_i(t) - S^{(2)}_i(t-1).
\label{eq:differentiation}
\end{equation}

\subsubsection{Dark subtraction}
The slow-scan observation mode took dark frames during the manoeuvre at the beginning and ending of each scan observation.
We calculate the dark current per pixel by taking the median value of all dark frame data, and subtract it from the pixel value, assuming that the dark level was almost constant in one scan,
\begin{equation}
S^{(4)}_i = S^{(3)}_i - D_i,
\label{eq:darksub}
\end{equation}
where $D$ is the dark signal.

\subsubsection{Flat fielding}
\begin{figure}
\begin{center}
\FigureFile(80mm,55mm){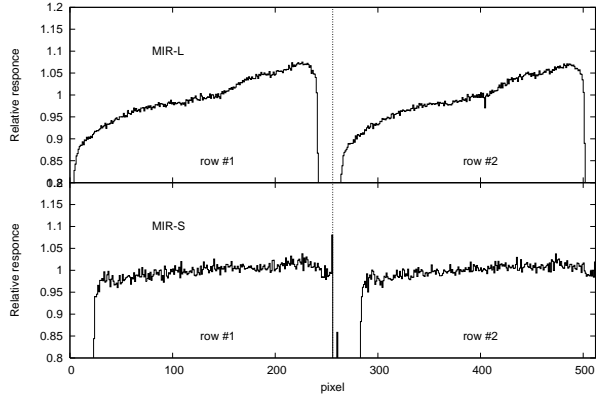}
\end{center}
\caption{
Flat data for the \textit{S9W} (bottom) and \textit{L18W} (top) bands 
normalized by the mean values of pixels 25--256 in row\#1 and 281--512 in row\#2 for the \textit{S9W} band 
and 2--236 in row\#1 and 258--492 in row\#2 for the \textit{L18W} band.
}
\label{fg:flat}
\end{figure}

We construct the flat-field data from $\sim$200 observations of the south ecliptic pole (SEP) region by taking the median value for each pixel.
The correction is applied as
\begin{equation}
S^{(5)}_i = S^{(4)}_i / F_i,
\label{eq:flatfield}
\end{equation}
where $F$ is the normalized flat correction factor (see Figure \ref{fg:flat}).

\subsubsection{Masking bad pixels}
We set flags on invalid data such as
\begin{enumerate}
\item pixels which were masked out for the slit spectroscopy,
\item data of the first 2 samples after each reset because of significant reset anomaly, and
\item saturated data which have ADU$>$40000 in the raw signal output.
\end{enumerate}
These data are not used to create images.

\subsection{Image creation}
After the basic calibrations, we proceed to image creation.
In this process, we use a GCF (Gridding Convolution Function; \cite{sawada2008}) to re-grid the sampling data onto a regular grid, since the data acquisition was spatially irregular in the sky plane.
The default GCF is a Gaussian function with full width at half maximum (FWHM) of 1 pixel, where the pixel sizes are 2\farcs34 and 2\farcs51 at the \textit{S9W} and \textit{L18W} bands, respectively.
To abandon charged-particle-hit data, we perform 5-$\sigma$ clipping within 1 pixel radius
\footnote{Usually there are more than 20 data in 1 pixel radius, although the number varies with the scan speed.}.
We note that threshold lower than 3-$\sigma$ will diminishes the peak intensity of a true source in some degree.

\subsection{Self-pointing reconstruction}
\label{sc:selfpr}

\begin{figure*}
\begin{center}
\FigureFile(160mm,50mm){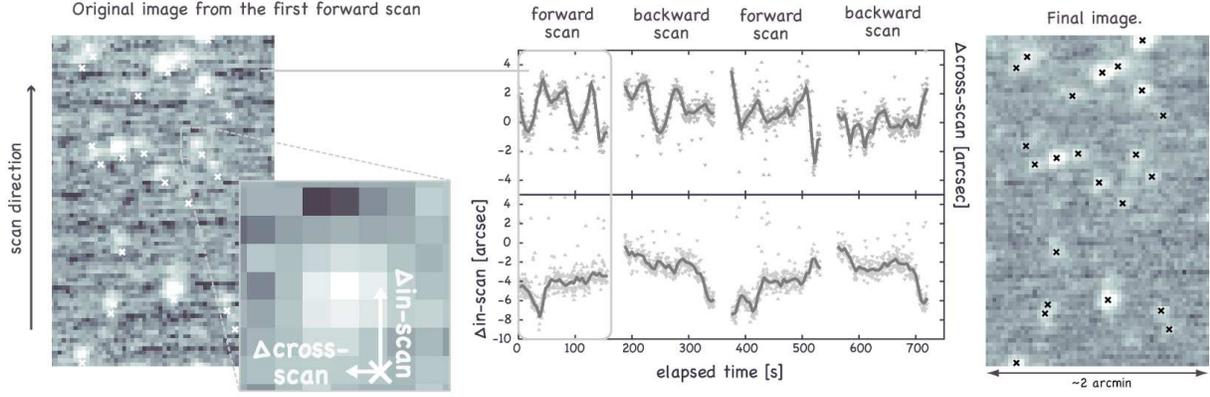}
\end{center}
\caption{
Example of the self-pointing reconstruction.
(\textit{left}) The 9 $\mu$m image from the first forward-scan using the satellite telemetry and a magnified image of a point source.
The pixel size is 2\farcs34.
The crosses are the positions of 2MASS point sources.
(\textit{middle}): Time variations of the positional differences along the cross-scan (top) and in-scan (bottom) directions, respectively.
The thick curves indicate the overall trends of the time variations of the positional differences.
(\textit{right}): The final image based on the data positionally corrected by the self-PR method.
}
\label{fg:selfpr}
\end{figure*}

In making images, we need positional information of every pixel at every sampling.
There are two kinds of information in TSD; AOCU (the Attitude and Orbit Control Unit) and G-ADS (the Ground-based Attitude Determination System) (see \cite{fis_dum}).
However, their positional accuracy is typically $10''$ in peak-to-peak, (see the left image in Figure \ref{fg:selfpr}).
Although these accuracies satisfy the requirement for the attitude controle system of AKARI of $30''$ absolute error, 
it is not accurate enough compared to the pixel size of 2\farcs5 and $5''$ in IRC51 and IRC11 observations, respectively.
Actually, there appear many fake double stars owing to the positional differences between forward and backward scans in each round trip.
Therefore, ARIS has a function to make time-dependent correction of scanned position based on the TSD by comparing the positions of detected point sources with those in the reference catalogs.
This process is called self-pointing reconstruction (self-PR).

The self-PR procedure is performed as follows.
\begin{enumerate}
\item We make two or four initial images of each individual one way scan in each round trip using either AOCU or G-ADS position data (see the left panel in Figure \ref{fg:selfpr}).
Usually, we use the \textit{S9W} data because we need as many point-like sources as possible in the images.
\item We extract point sources and store their observed time and positions.
To avoid extracting fake sources, we use a high threshold level of 5-$\sigma$.
\item We do cross-identification between the detected sources in our images and the point sources listed in the reference catalogs.
As a reference catalog, we usually use the Two-Micron All Sky Survey (2MASS) Point Source Catalog, whose positional accuracy is as good as 1$''$.
Since the 2MASS sensitivity is as good as 14~mag in the $K_S$ band, most of the AKARI sources are easily identified in the catalog.
\item We evaluate the differences between our positions and those in the catalog.
Here, the positional differences are measured in the in-scan and cross-scan directions.
\item We take mean values of the differences in a given time interval (30~sec as a default value), and then, connect the values by line segments to get the overall trend (see the middle panel in Figure \ref{fg:selfpr}).
The computed trends are output as a text file as time vs. positional difference format.
\item We revise the position information in TSD with the computed trends and make the final image from all forward/backward scans.
\end{enumerate}

The positional uncertainty in the final image is as small as 1$''$ (see the right image in Figure \ref{fg:selfpr}).
In addition, we can also correct the positions for the \textit{L18W} and the FIS slow-scan data, which were observed simultaneously, using the trend file made from the \textit{S9W} data.
The application to the FIS data was demonstrated in \citet{ikeda2012}.
We note that the performance of the self-PR method becomes maximum when sufficient number of bright \textit{S9W} sources are distributed over the scanned area.

\section{Flux Calibration for Point Sources}
\label{sc:aris_cal}

\subsection{Observed standard stars and data processing}
For the absolute flux calibration, we observed 4 and 3 infrared standard stars \citep{cohen1999} for the \textit{S9W} and \textit{L18W} bands, respectively (Table \ref{tbl:standard}).
These observations were carried out in the AOT IRC51.

The data were processed by ARIS, and the final images were created using the self-PR method.
We applied aperture photometry to the final images.
We used the DAOPHOT-Type Photometry Procedures of GCNTRD and APER in the IDL Astronomy User's Library (version 2007 May; \cite{landsman1993}).
The radius of the aperture was 7.5 pixels and the sky level was estimated from the annulus between 7.5 and 12.5 pixel radii.

\begin{table*}
\caption{Observed standard stars.}
\label{tbl:standard}
\begin{center}
\begin{tabular}{lllllll}
\hline
 & & \multicolumn{2}{c}{Expected flux density [Jy]$^{\dagger}$} & \multicolumn{3}{c}{Observation}  \\
\cline{3-4}
\cline{5-7}
Star & Sp. type & \textit{S9W} & \textit{L18W} & date (UT) & ID & param$^{\ddagger}$  \\
\hline
BD+62 1588 & K5 III & 4.9966$\times10^{-1}$ & 1.1341$\times10^{-1}$ & 2007-03-13 23:59:21 & 5124045 1 & i;N;15 \\
                               &        &            &            & 2007-03-14 01:38:43 & 5124046 1 & i;N;30 \\
                               &        &            &            & 2007-03-16 00:00:36 & 5124047 1 & i;L;15 \\
                               &        &            &            & 2007-03-16 01:39:57 & 5124048 1 & i;L;30 \\
\hline
HD 42525 & A0 V & 2.7792$\times10^{-1}$ & 5.8763$\times10^{-2}$ & 2007-05-02 09:00:33 & 5124066 1 & i;N;30 \\
                             &      &            &            & 2007-05-03 01:34:32 & 5124065 1 & i;N;15 \\
                             &      &            &            & 2007-05-04 23:57:43 & 5124067 1 & i;L;15 \\
                             &      &            &            & 2007-05-05 01:37:07 & 5124068 1 & i;L;30 \\
\hline
HD 46819 & K0 III & 4.4501$\times10^{-1}$ & 9.6344$\times10^{-2}$ & 2007-06-06 00:46:43 & 5124108 1 & a;N;15 \\
                             &        &            &            & 2007-06-06 02:26:08 & 5124109 1 & a;N;30 \\
                             &        &            &            & 2007-06-17 01:56:29 & 5124110 1 & i;L;15 \\
                             &        &            &            & 2007-06-20 02:51:59 & 5124111 1 & i;L;30 \\
\hline
BD+66 1060 & K2 III & 1.1766$\times10^{-1}$ & 2.6382$\times10^{-2}$ & 2007-08-02 04:29:25 & 5124130 1 & a;N;30 \\
\hline
\end{tabular}
\end{center}
$\dagger$ Flux density for each band estimated from equation (\ref{eq:in-band}).\\
$\ddagger$ The first characters `a' and `i' indicate one and two round trips, respectively.
The second parameter shows that the target source is observed with the MIR-S (`N') or MIR-L (`L') channel.
The last numeral is the scan speed in arcsec~sec$^{-1}$.
\end{table*}

\subsection{Estimation of the in-band flux density}
The in-band flux density of each band at the effective wavelength, $f^{\rm quoted}_\lambda (\lambda_i)$ is calculated by the following equation:
\begin{equation}
f^{\rm quoted}_\lambda (\lambda_i) =
\frac{\int_{\lambda_{is}}^{\lambda_{ie}} R_i(\lambda) \lambda f_\lambda(\lambda) d\lambda}
{\int_{\lambda_{is}}^{\lambda_{ie}} (\lambda_i/\lambda) R_i(\lambda) \lambda d\lambda},
\label{eq:in-band}
\end{equation}
where $f_\lambda(\lambda)$ is the Spectral Energy Distribution of a standard star (provided by M. Cohen) and $R_i(\lambda)$ is the spectral response function (including the transmission of the optics and the response of the detector) of the band $i$.
The in-band flux density becomes equal to a true value, only if $f_\lambda \propto \lambda^{-1}$.
This is the convention adopted by IRAS, COBE, ISO, Spitzer/IRAC, and the other AKARI observation modes \citep{tanabe2008,ishihara2010}.

\subsection{Conversion factor}
\begin{figure*}
\begin{center}
\FigureFile(80mm,80mm){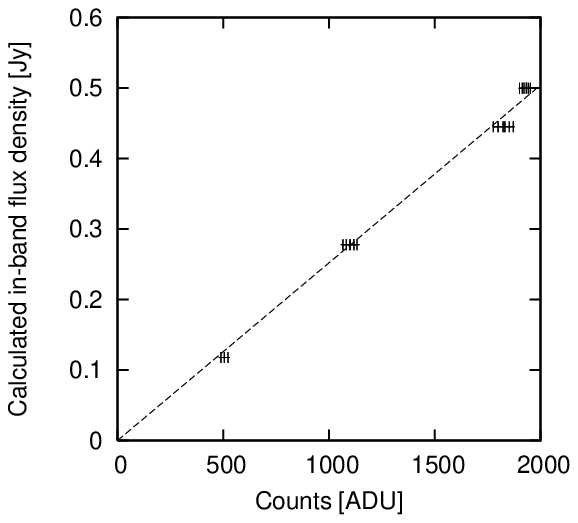}
\FigureFile(80mm,80mm){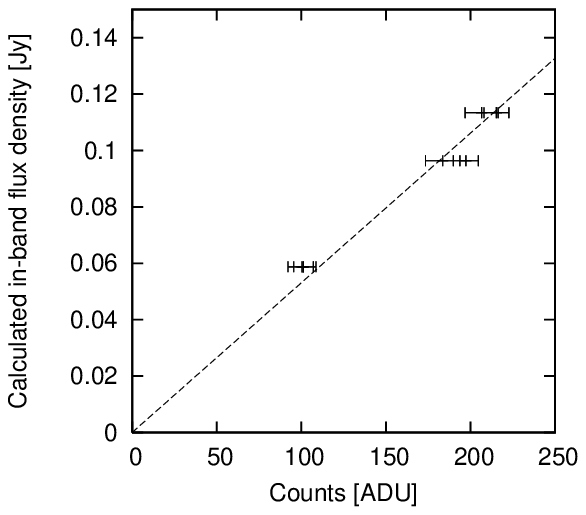}
\end{center}
\caption{
Estimated flux density versus observed ADU of the standard stars for the \textit{S9W} (\textit{left}) and \textit{L18W} (\textit{right}) bands.
A broken line in each plot represents the least-square fit to the data.
}
\label{fg:cfactor}
\end{figure*}

The relations between the intensities in ADU from the images and the calculated in-band flux densities in Jy of the standard stars are shown in Figure \ref{fg:cfactor}.
Straight lines were fitted to these data and the slopes of the fitted lines provide the conversion factors $cf$ in Jy~ADU$^{-1}$, as listed in Table \ref{tbl:cfactor}.

We note that the calibrations for diffuse emission have not been established yet.

\begin{table}
\caption{Conversion factor $cf$ [Jy~ADU$^{-1}$]}
\label{tbl:cfactor}
\begin{center}
\begin{tabular}{lcc}
\hline
Band & $cf$ & Error\\
\hline
\textit{S9W}  & $2.518 \times 10^{-4}$ & $3.228 \times 10^{-6}$\\
\textit{L18W} & $5.291 \times 10^{-4}$ & $1.365 \times 10^{-6}$\\
\hline
\end{tabular}
\end{center}
\end{table}

\section{Performance of IRC Slow-scan Observations}

We evaluate the performances of the IRC slow-scan observations by detected point sources, such as detection limits, positional accuracies, and flux uncertainties, on the basis of the 146 pointed observations toward the Chamaeleon region ($>$30~deg$^2$).

\begin{figure}
\begin{center}
\FigureFile(80mm,80mm){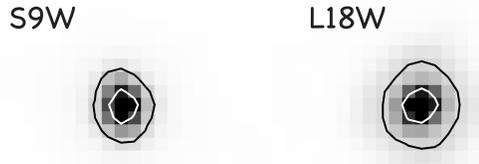}
\end{center}
\caption{
PSF images of the \textit{S9W} (\textit{left}) and \textit{L18W} (\textit{right}) bands.
The pixel scale is 2\farcs34 and 2\farcs51 for the \textit{S9W} and \textit{L18W}, respectively.
The black and white contours in each panel represent the 10~\% and 50~\% levels of the peak value, respectively.
}
\label{fg:psf}
\end{figure}

At first, we constructed images of the \textit{S9W} and \textit{L18W} bands using ARIS, applying the self-PR method.
We have succeeded in extracting 2974 and 492 point-like sources from the \textit{S9W} and \textit{L18W} images, respectively, and 246 sources were detected in both the bands.
We created point spread functions (PSFs) with 25 brightest and relatively isolated stars, excluding saturated ones, in the \textit{S9W} and \textit{L18W} images toward the Chamaeleon I molecular cloud using the `GETPSF' procedure in the IDL DAOPHOTO package.
Figure \ref{fg:psf} shows the derived PSFs; the FWHM sizes of the PSFs are $\sim6''$ and $\sim7''$ for the \textit{S9W} and \textit{L18W} bands, respectively.

\subsection{Sensitivity}

\begin{figure}
\begin{center}
\FigureFile(80mm,56mm){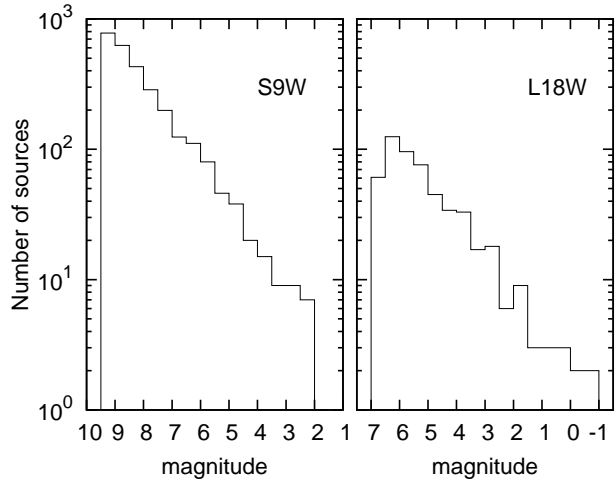}
\end{center}
\caption{
Histograms of the magnitudes of the extracted sources in the \textit{S9W} (\textit{left}) and \textit{L18W} (\textit{right}) bands.
}
\label{fg:lf}
\end{figure}

Figure \ref{fg:lf} shows histograms of the extracted sources by the Chamaeleon observations.
The lower limit in \textit{S9W} and the turnover in \textit{L18W} are thought to correspond to the detection limits of the AKARI IRC slow-scan observations; 9.5~mag for the \textit{S9W} band and 6.5~mag for \textit{L18W}.
These limits are superior to those of the AKARI All-Sky Survey of 0.05~Jy (7.6~mag) and 0.09~Jy (5.3~mag) at the \textit{S9W} and \textit{L18W} bands, respectively \citep{ishihara2010}.
Note that the sources brighter than 2 mag at the \textit{S9W} band are thought to be saturated.

\subsection{Positional accuracy}

\begin{figure}
\begin{center}
\FigureFile(80mm,56mm){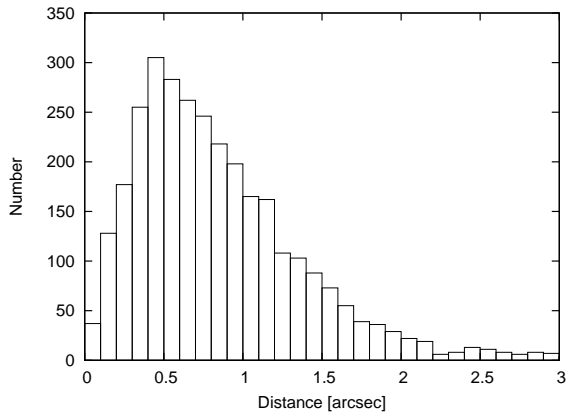}
\end{center}
\caption{
Histograms of the angular distances between the AKARI \textit{S9W} and 2MASS sources.
}
\label{fg:position}
\end{figure}

We checked the positional accuracy of the detected sources by comparing with the 2MASS Point Source Catalog.
We searched 2MASS counterparts with positional difference of 3$''$ (see \cite{ita2010}) and found that 3074 out of 3220 AKARI Chamaeleon sources have counterparts.
Figure \ref{fg:position} shows the positional differences between our sources and the counterparts.
The mean and standard deviation values of the differences are calculated as $0\farcs87 \pm 0\farcs58$, which is negligible for the full IRC resolution of 2\farcs5.
We note that we cannot see any correlations between positional differences and flux densities.
There are 146 sources (126 for \textit{S9W} and 22 for \textit{L18W}) without 2MASS counterparts.
Since about 90 \textit{S9W} sources without 2MASS counterparts are as faint as $>$8.5~mag, we cannot exclude the possibility that these faint sources are fake ones considering the detection limit of 9.5~mag at the \textit{S9W} band.
Furthermore, since most of other sources are located near the edge of the images, they are also thought to be fake ones.
On the other hand, the absence of 2MASS counterparts for 22 \textit{L18W} sources were thought to be caused by the limitation of the self-PR method;
since the self-PR method was applied using bright \textit{S9W} sources, we could not improve the positional accuracy for regions where no bright \textit{S9W} sources were detected (see \S \ref{sc:selfpr}).
Indeed, 16 out of 22 sources have 2MASS counterparts within a positional difference of 5$''$.

\subsection{Flux uncertainty}

\begin{figure*}
\begin{center}
\FigureFile(80mm,40mm){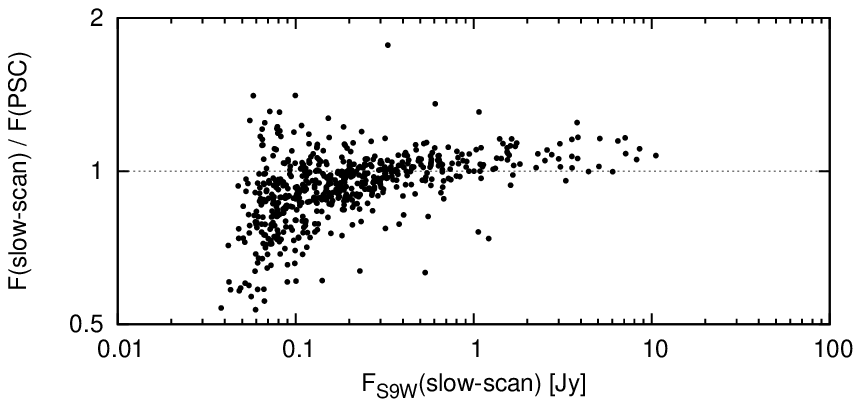}
\FigureFile(80mm,40mm){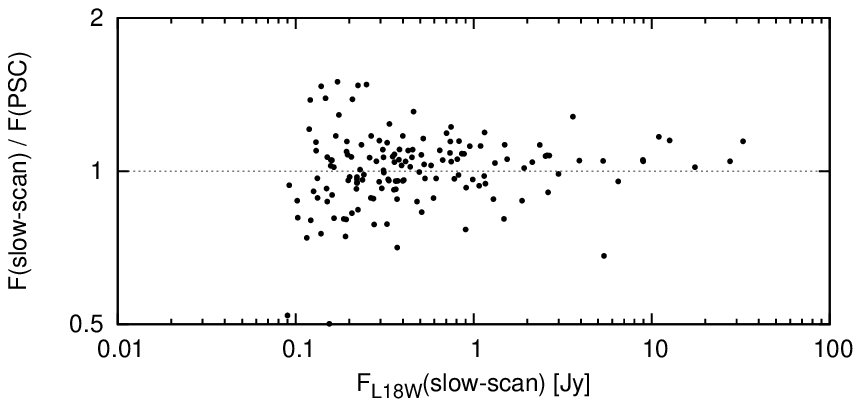}
\end{center}
\caption{
Flux density ratios of the slow-scan to the IRC PSC as a function of the slow-scan flux densities for the \textit{S9W} (\textit{left}) and \textit{L18W} (\textit{right}) bands.
}
\label{fg:flux_comp}
\end{figure*}

We compared the flux densities of the detected sources with those in the AKARI IRC PSC to verify the flux calibration of the IRC slow-scan observations.
There were 613 and 150 sources detected at both the slow-scan and All-Sky Survey in the \textit{S9W} and \textit{L18W} bands, respectively.
Figure \ref{fg:flux_comp} shows the ratio of the slow-scan to PSC flux densities as a function of the slow-scan flux density.
It seems that the ratios for the faint ($<$0.1~Jy) \textit{S9W} sources drops systematically down to $\sim$0.5--0.7.
This disagreement is thought to be caused by the `flux-boost' effect in the IRC PSC.
It is explained as follows;
if a source is as faint as the noise level, we can detect it only when the signal is unevenly bright.
Actually, the completeness of the IRC PSC is only $\sim$50~\% at $f(9\micron) = 0.1$~[Jy] (see Figure 20 and Table 3 of \cite{ishihara2010}).
The mean values and standard deviations of the ratios are listed in Table \ref{tbl:flux_ratio}.

\begin{table}
\caption{Mean values (MEAN) and standard deviations (STDDEV) of the flux ratio of the slow-scan to the All-Sky Survey.}
\label{tbl:flux_ratio}
\begin{center}
\begin{tabular}{lccc}
\hline
Band & MEAN & STDDEV & n$_{\rm star}^\star$\\
\hline
\textit{S9W}  & $0.945$ & $0.178$ & 613\\
($>0.1$[Jy])  & $0.986$ & $0.160$ & 423\\ \hline
\textit{L18W} & $1.027$ & $0.195$ & 150\\
\hline
\end{tabular}
\end{center}
$\star$ `n\_star' indicates the number of the stars used in the estimation.
\end{table}

\begin{figure*}
\begin{center}
\FigureFile(80mm,56mm){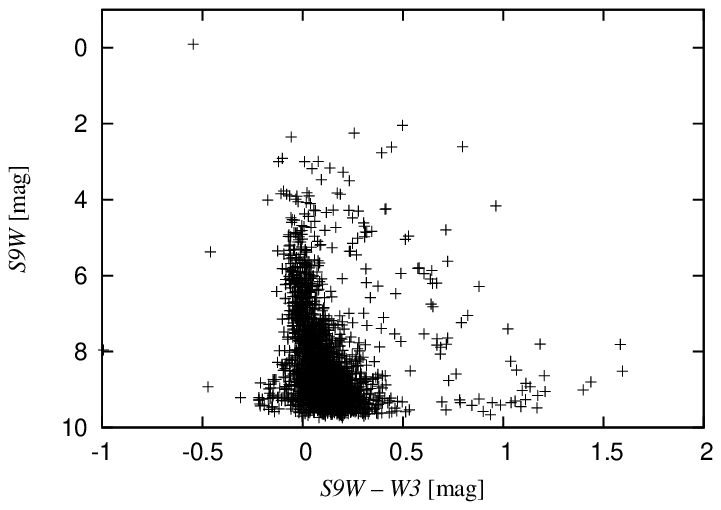}
\FigureFile(80mm,56mm){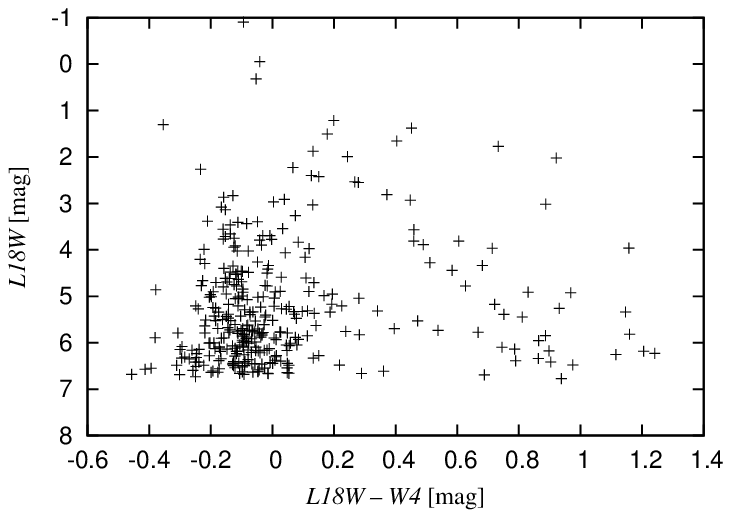}
\end{center}
\caption{
Color-magnitude diagrams of \textit{S9W}-\textit{W3} (WISE 12 $\mu$m) vs. \textit{S9W} (\textit{left}) and \textit{L18W}-\textit{W4} (WISE 22 $\mu$m) vs. \textit{L18W} (\textit{right}).
}
\label{fg:cmd}
\end{figure*}

We also searched for the counterparts in the first preliminary release of the WISE (the Wide-field Infrared Survey Explorer; \cite{wright2010}) catalog to verify AKARI flux densities.
WISE surveyed the entire sky at 3.4, 4.6, 12 and 22 $\mu$m bands (which are referred to as W1, W2, W3, and W4) with spatial resolutions of $6''$ at W1--W3 and $12''$ at W4.
The WISE preliminary catalog covers our surveyed region.
We found that 3088 out of 3220 sources have WISE counterparts within 6$''$ searching radii.
Here, we considered only the WISE sources with SNR $>$ 10 in the 12 $\mu$m bands.
Figure \ref{fg:cmd} shows the \textit{S9W}-\textit{W3} vs. \textit{S9W} and the \textit{L18W}-\textit{W4} vs. \textit{L18W} color-magnitude diagrams.
In the diagrams, excess sources such as young stellar objects, asymptotic giant branch stars, and galaxies are located on the right-hand side.
For most sources, AKARI and WISE measurements are in good agreement to each other considering the differences in the band passes;
 the mean colors are 0.12 and 0.037 at \textit{S9W}-\textit{W3} and \textit{L18W}-\textit{W4}, respectively.

\section{Summary}

We have developed a toolkit, ARIS (AKARI data Reduction tools for IRC Slow-scan), and described in detail the data reduction processes.
Using the self-PR method in ARIS, the positional accuracy of final images reaches as good as $1''$, smaller than the pixel size of IRC of 2\farcs5.
The PSF sizes are $\sim6''$ and $\sim7''$ in FWHM for the \textit{S9W} and \textit{L18W} bands, respectively.
We also performed flux calibrations using infrared standard stars.
We further checked the uncertainties of flux densities comparing with the AKARI IRC PSC and the WISE preliminary catalog, and found that the uncertainties are less than 20~\%.

This work is based on observations with AKARI, a JAXA project with the participants of ESA.
We gratefully acknowledge F.~Usui, C.~Yamauchi and all the members of the AKARI project for their support to this study.
This study uses data products from the Two Micron All Sky Survey, which is a joint project of the University of Massachusetts and the Infrared Processing and Analysis Center/California Institute of Technology, funded by the National Aeronautics and Space Administration and the National Science Foundation.
WISE is a project of University of California, Los Angeles, and Jet Propulsion Laboratory (JPL)/California Institute of Technology (Caltech), funded by the National Aeronautics and Space Administration (NASA).
This research has made use of the SIMBAD database and the VizieR catalog access tool, CDS, Strasbourg, France.
S.T. is financially supported by the Japan Society for the Promotion of Science.

\end{document}